\documentclass{article}

\usepackage{agents}
\usepackage[bf,hang]{subfigure}
\usepackage{graphicx}
\usepackage[bf,hang]{caption}

\title{Adaptivity in Agent-Based Routing for Data Networks}

\author{
David H. Wolpert\\
NASA Ames Research Center\\
{\tt dhw@ptolemy.arc.nasa.gov} 
\and
Sergey Kirshner\\
University of California, Irvine\\
{\tt skirshne@ics.uci.edu} \\
\and
Chris J. Merz \\
NASA Ames Research Center\\
Caelum Research \\
{\tt cmerz@ptolemy.arc.nasa.gov}\\
\and
Kagan Tumer\\
NASA Ames Research Center\\
{\tt kagan@ptolemy.arc.nasa.gov}\\ [.2in]
Tech Rep NASA-ARC-IC-1999-122
}

\begin{document}

\maketitle

\begin{abstract}

Adaptivity, both of the individual agents and of the interaction
structure among the agents, seems indispensable for scaling up
multi-agent systems (MAS's) in noisy environments.  One important
consideration in designing adaptive agents is choosing their action
spaces to be as amenable as possible to machine learning techniques,
especially to reinforcement learning (RL) techniques ~\cite{suba98}. 
One important way to have the interaction structure
connecting agents itself be adaptive is to have the intentions and/or
actions of the agents be in the input spaces of the other agents, much
as in Stackelberg games~\cite{baol99,kola95,kola97,laor97}. 
We consider both kinds of adaptivity in the design of a MAS
to control network packet routing~\cite{sudr97,boli94,kumi97,hesn98}
We demonstrate on the OPNET
event-driven network simulator the perhaps surprising fact that simply
changing the action space of the agents to be better suited to RL can
result in very large improvements in their potential performance: at
their best settings, our learning-amenable router agents achieve
throughputs up to three and one half times better than that of the standard
Bellman-Ford routing algorithm, even when the Bellman-Ford protocol
traffic is maintained. We then demonstrate that much of that potential
improvement can be realized by having the agents learn their
settings when the agent interaction structure is itself adaptive.

\end{abstract}

\section{Introduction}

As time goes on, larger and larger multi-agent systems (MAS's) are
being deployed as a way to meet a single overall goal for an
underlying system, and this is being done for increasingly noisy and
unreliable systems~\cite{clbo98,goro99,huwe98b,jesy98,sale97,sen97,syca98}. 
However if one uses
traditional ``hand-tailoring'' to design all aspects of a MAS,
maintaining robustness while scaling up to large problems becomes
increasingly difficult. Accordingly, it is becoming imperative to
understand how best to have both the individual agents and the
structure of their interactions be as adaptive as possible.

In designing agents to be adaptive one should cast their action spaces
in a form that is as amenable as possible to machine learning
techniques, especially to reinforcement learning (RL) techniques
~\cite{suba98,wada92}. However it is often the case that more than
just the policies of the individual agents needs to be adaptive; for
the system to perform well, often the very structure with which the
agents interact also needs to be adaptive rather than hard-wired
~\cite{wotu99b,wotu99a,wowh99a}.  
One way to have that structure be adaptive is to
exploit the existence of input spaces in RL-based agents by having the
intentions and actions of the agents be in the input spaces of the
other agents, much as in Stackelberg games~\cite{baol99,kola97,kola95,laor97}.
In this way as individual agents adapt their
policies, information concerning the best way to adapt to those new
policies is automatically propagated to the other agents.

We consider both kinds of adaptivity in the design of a MAS to control
network packet routing with the goal of maximizing
throughput~\cite{boli94,hesn98,kumi97,sudr97}.  
In this domain the naive
choice of the action space of each router agent is the categorical
variable of the single outbound link along which to route the
particular packet currently at the top of its queue. However in
comparison to Euclidean variables with their inherent smoothness
structure, such a categorical variable is usually poorly suited to RL
techniques. Accordingly, we instead consider having the action space
be the vector of the proportion of packets the agent routes along each
of its outbound links. Packets can be routed according to such a
proportion vector by taking that vector to specify probabilities for
all routing along all outbound links. Alternatively, one can route
traffic deterministically, in such a way that the proportions of the
traffic actually sent are as close as possible, according to any of a
suite of metrics, to the desired proportion vector. In this paper we
concentrate on the second of these schemes.\footnote{We have
developed a particularly fast implementation of this second scheme, an
implementation that can also have built-in ``data-aging'', so that
more recent traffic is counted more heavily than older traffic. In
addition, this second scheme can be modified to avoid
``round-robining'', so that packets do not arrive out of sequence at
their ultimate destination. See ~\cite{wolp99a} for this and
other extensions of this scheme.} Whichever proportion-vector-based
scheme one uses, one can have each agent learn how best to set the
vectors it uses (one vector for each potential ultimate destination of
the packet), and in this way adaptively determine how best to do its
routing.

One difficulty with this new action space is that it easily results in
``cycles'', in which a particular packet may return to an agent that
had previously routed it. To avoid this, we developed the {\bf hard
masking} routing algorithm, which maps an original proportion vector
to a new one. This algorithm has the property that if it is used by
every agent, then any links leaving an agent that could result in
cycles are ``masked out'', so that no traffic is sent along that
link. (In fact, in our experiments even if some agents do not use
hard masking, if they employ conventional shortest-path routing
algorithms, cycles will still be avoided.) At the same time, the
ratios of traffic sent along all non-masked links are left
unchanged. Hard masking requires no additional protocol traffic beyond
that already contained in traditional distance-vector or link-state
routing algorithms ~\cite{ahma93,bega92}. However if the agents use RL
to learn proportion vectors that are run through the hard masking
algorithm before being used, the system potentially can adapt to use
far better proportions than those that arise (in a completely
unintended manner) when one uses traditional algorithms, while sharing
with those algorithms the absence of any risk of cycling.

Unfortunately, hard masking has a clipping property: it does not affect
the amount of traffic being sent down a link until a certain property
of that link reaches a threshold, at which point all traffic down that
link is blocked. As one might expect, this hard clipping can reduce
routing efficacy. To overcome this problem, we developed the {\bf soft
masking} routing algorithm, which gradually decreases traffic along a
link as the threshold is approached, while still preventing
cycles. The version of soft masking we use is optimal in that it is
the unique variant of hard masking that preserves invariance both in
rescaling of time and/or packet sizes, and in translation of the
zero-point of one's clock.

We first investigated hard and soft masking on the OPNET event-driven
network simulator, without any learning on the part of the agents; we
simply swept through the space of potential proportion vectors,
recording performance as we went. These experiments were on relatively
simple and small networks (currently all TCP/IP-based networks are
broken down for routing purposes into subnetworks almost always having
no more than a dozen routers).  These runs demonstrated that at the
optimal proportion vectors, masking can result in throughput up to
five times better than that of Bellman-Ford (BF), the traditional routing
algorithm we used as our comparison point. This improvement was
achieved even though the full BF routing protocol traffic
was still running ``in the background'' in the masking systems. We
also found that the size of the basins in the space of potential
proportion vectors which gave at least some improvement over
BF was quite large --- approximately half of the range of
each component of each proportion vector in the case of soft masking.

As mentioned above, the second component of an adaptive MAS beyond
having the individual agents be adaptive is having the agent
interaction structure itself be adaptive, so that the agents can
automatically and adaptively cooperate with one another. To that end,
we considered two possible interaction structures. The first can be viewed as an
iterative Stackelberg game structure~\cite{baol99,kola97,kola95,laor97}.
In this structure, ``leader'' agents first
determine what proportion vectors they will use, and then the
``follower'' agents use that information to determine what proportion
vectors they think will result in optimal performance. In other words,
the proportion vectors of the leaders are components of the input
space of the followers. We also investigated a less asymmetric
structure, in which agents ``interleaved'' their decisions in such a
way so that every agent was in some respects acting as a follower and in
some respects acting as a leader. 

We investigated how much of the potential improvement of masking over
BF can be realized by having the agents learn their proportion vectors
using these kinds of adaptive agent interaction structures. We found
that even using extremely simple RL algorithms, and with essentially
no effort given to optimizing the soft masking, when those agents
operated within the adaptive structure outlined above, typically
throughput was 3 times better than with BF. We never encountered
an instance in which soft masking consistently gave worse performance.

In Section~\ref{sec:routing} we describe conventional routing algorithms,
proportional routing, and the various forms of masking. 
In Section~\ref{sec:learn} we describe the learning schemes used and the 
adaptive agent interaction schemes investigated. In Section~\ref{sec:results} 
we present the results of our experiments.

\section{Agents for Network Routing}
\label{sec:routing} 
\subsection{Shortest Path Routing}
%

The most commonly used routing algorithms are based on the ``shortest
path'', i.e., the path from a router to a destination that would experience
the minimal cost if the traffic were routed down that path. In such
algorithms each router stores the smallest of all possible costs to
each destination, along with the first link on the associated
path. (This data is commonly stored, sometimes along with other
information, in a ``routing table.'')  The router then sends all its
packets bound for a particular destination along the first link on the
associated shortest path.  There are many algorithms for efficiently
computing the shortest paths when the costs for traversing each router
and link in the network are known, including Dijkstra's
algorithm~\cite{ahma93,bega92,depa84,dijk59} and the BF
algorithm~\cite{bell57,bell58,bega92,fofu56}. When applied in dynamic
data networks of the kinds considered here, both algorithms entail
some underlying protocol traffic, to allow the routing tables of the
routers to adapt to changes in traffic conditions.


\subsection{Proportional Routing Agents}

Shortest path algorithms in general and BF in particular have several
shortcomings.  In practice, the shortest path estimates are always
based on old information, which means each router bases its routing
decisions on potentially incorrect assumptions about the network.
However even if a shortest path algorithm is provided the exact
current costs of all the links, because it sends $all$ of its traffic
with the same ultimate destination down a single link, such an
algorithm still provides suboptimal solutions. (Formally, this
suboptimality holds so long as we're not in the limit where each
router makes infinitesimal routing decisions at each moment, with its
routing table being updated before the next infinitesimal routing
decision --- see ~\cite{boli94,kumi97,bega92,tuwo99}.)

This second problem with BF can potentially be alleviated if each
routing agent apportions its traffic bound for a particular
destination along more than one path, rather than sending it all along
the shortest path. In this paper we are concerned with agents that
learn, dynamically, how best to do this. As discussed above, we are
interested in having each agent do its learning with an action space
that consists of one proportion vector $\vec{p}$ satisfying:
$$
0\leq p_i\leq 1, \; \; i=1,\ldots,m \;\;\;\; \mbox{and} \;\;\;\;
\sum_{i=1}^m p_i=1
$$ 
for each destination, $m$ being the number of
outbound links. This proportion vector then determines how the traffic
bound to that destination from that router gets apportioned among that
router's outbound links, as discussed above. We call this
``proportional routing''.


\subsection{Hard Masking}

Simple proportional routing invariably results in unproductive cycles
being introduced into the paths followed by some packets.  One way to
avoid such cycles employs a destination-dependent ordering $v(r)$
over all routing agents. Given such an ordering, we can restrict
router $r_1$ to only send out packets according to its proportion
vector along those links connecting $r_1$ to routers $r_i$ such that
$v(r_i) \; < v(r_1)$; no traffic is sent along any other
link. Assuming all routers have the same ordering $v(r)$ for the same
destination, having them all follow this scheme ensures that there
will be no cycles. (In our work, for convenience, we choose $v(r)$ for
each destination $d$ to be the smallest cost estimates for going from
$r$ to $d$ stored in the routing table on $r$.)

We use the term ``masking'' to refer to any scheme of this nature in
which the components of $\vec{p}$ are multiplied by constants set by
the condition of the network. In particular, to define hard masking,
let our routing agent be $r_1$, let the destination be $d$, let the
router neighbors of $r_1$ be the $\{r_i\}$, let $r_1$'s proportion
vector be $\vec{p}$, and let the ordering over routers for destination
$d$ be $v(r)$. Then in the technique of hard masking we calculate an
{\bf applied proportion vector} $\vec{p'}$ from $\vec{p}$ according to the
following formula:
\begin{equation}
p'_i \equiv \frac{p_i \;\; \Theta(v(r_1) - v(r_i))}
{\sum_j {p_j \;\; \Theta(v(r_1) - v(r_i))} } \; ,
\end{equation}

\noindent
where $\Theta(x)$ is the Heaviside function that equals 1 for positive
argument, and equals 0 otherwise. We then use $\vec{p}'$ rather than $\vec{p}$ to
govern the routing from router $r_1$. (From now on, when we need to
distinguish it from $\vec{p'}$, we refer to
$\vec{p}$ as a {\bf base proportion vector}.)


\subsection{Soft Masking}

Although hard masking does avoid cycles while still having the generic
behavior of not sending all traffic bound for a particular destination
down a single link, it does so in a potentially brittle manner. This
is because a link will either be used fully (according to the
proportion vector), or, for what may only be an infinitesimal change
in network conditions, not used at all. A more reasonable strategy
would be for the routing agent to only gradually reduce its traffic
along any link $i$ as that link approaches the condition $v(r_1) =
v(r_i)$, in such a way that $p'_i = 0$ when $v(r_1) = v(r_i)$.  If it
does this, a routing agent $r_1$ will have essentially replaced hard
masking's discontinuous {\bf masking function} $M_{r_1}(v(r_1),
v(r_i)) \equiv \Theta(v(r_1) - v(r_i))$(i.e., the function that gets
multiplied by $p_i$ in the determination of the applied vector
$p'_i$) with a continuous one.

What is the best way to implement such a ``gradual reduction of
traffic''?  One obvious requirement is that the new masking function
be both translation and scaling invariant with respect to changes to
the functions $v(.)$, since those functions only provide an
ordering. In particular, in our implementation where the $v(r)$ are
costs given by times, we don't want either the zero-points or the
units with which we measure time to matter --- changes to either
should not affect the behavior of the router.

To ensure translation invariance, it suffices to require that
$M_{r_1}(v(r_1), v(r_i))$ be of the form $ M_{r_1}(v(r_i)-v(r_1))$.
For scaling invariance, we need to have the function $M_{r_1}$ obey
the following condition:
\begin{eqnarray*}
\frac{M_{r_1}(x)}{M_{r_1}(y)} =
\frac{M_{r_1}(a x)}{M_{r_1}(a y)} 
\end{eqnarray*}
for any $a \neq 0$. In other words, to preserve the ratios of traffic
sent along all links  under rescaling, for any values $x$ and $y$ the ratio
$\frac{M_{r_1}(a x)}{M_{r_1}(a y)}$ needs to be a constant,
independent of $a$. 

Now to make sure that no traffic is sent down a link once $v(r_i) \ge
v(r_1)$, write $M_{r_1}(x) \equiv N_{r_1}(x) \Theta(x)$. (Note that
$\Theta(v(r_1) - v(r_i))$ is both translation and scaling invariant.)
Restricting ourselves to the regime where $x > 0$ so that $N_{r_1}(x)
= M_{r_1}(x)$ and differentiating both sides of the scaling invariance
condition with respect to $a$ yields

\begin{eqnarray*}
x \frac{N'_{t}(a x)}{N_{t}(a x)} \; = \; y \frac{N'_{t}(a y)}{N_{t}(a y)}, 
\end{eqnarray*}
which must hold for any $a$, $x$ and $y$. In particular, take $a=1$,
and fix $x$, to get the following:
\begin{eqnarray}
y \frac{N'_{t}(y)}{N_{t}(y)} \; = A \;\;\;\; 
\mbox{where $A$ is  a constant.} 
\label{eq:A}
\end{eqnarray}
Now define $T_{r_1}(y) \equiv ln[N_{r_1}(y)]$. Having done this, equation~\ref{eq:A} becomes
$T'_{r_1}(y) = A/y$.  Integrating both sides, we get
\begin{eqnarray}
T_{r_1}(y) = D ln(y) + E 
\label{eq:D}
\end{eqnarray}

\noindent
where $D$ and $E$ are constants.
Exponentiating both sides, and recalling that $T_{r_1}(y) = ln[N_{r_1}(y)]$,
we get the solution
\begin{eqnarray}
N_{r_1}(x) = x^\beta \; .
\label{eq:M}
\end{eqnarray}

\noindent
(The overall multiplicative constant has been set to 1; it is
irrelevant in that it gets divided out when one divides by the
normalizing factor to calculate $\vec{p}$.)

Combining the two invariance properties gives us the final
soft masking function:
\begin{eqnarray}
N_{r_1}(x, y) = (x-y)^\beta.
\label{eq:M2}
\end{eqnarray}
So for routing agent $r_1$, ``soft masking'' means that the applied
proportion vector is set by the following:
\begin{equation}
p'_i \equiv \frac{p_i \;\; \Theta\left(v(r_1)-v(r_i)\right) \;\;
	e^{\beta\left(v(r_1)-v(r_i)\right)}}
{\sum_j p_j \;\; \Theta\left(v(r_1)-v(r_j)\right) \;\;
	e^{\beta\left(v(r_1)-v(r_j)\right)}} \; .
\end{equation}


\subsection{Implementation of Proportional Routing}

Perhaps the most straight-forward implementation of proportional
routing is for each routing agent to use a random number generator
with probabilities set to the proportion vectors to decide where to
route each successive packet.  This simple scheme has a major drawback
however. For large numbers of packets the realized proportions of the
packets actually sent will approximate the actual proportions
arbitrarily well. However this is not the case when the number of
packets is small. In particular, when masking is used, both the actual
proportion vectors (as formally defined above) and the actually
realized routing proportions will tend to change fairly frequently.  A
probabilistic approach may not result in such changing proportions
tracking each other accurately.

\begin{table*} [htb] \centering
\caption{Deterministic Proportional Routing (all entries given  
for $i \; \in {1,2,3}$)}
\vspace{.1in}
\begin{tabular}{c|c|c|c|c}
\hline \# sent & \# sent via & \# desired via & differences & chosen\\ 
& each route & each route & & link\\ 
$pkt_{total}$ & $pkt_i$ & $(pkt_{total}+1)*p_i$ &
$(pkt_{total}+1)*p_i-pkt_i$ & $l_i$\\ \hline \hline 
0 & (0, 0, 0) & (0.59, 0.31, 0.1) & (0.59, 0.31, 0.1)   & $l_1$ \\ 
1 & (1, 0, 0) & (1.18, 0.62, 0.2) & (0.18, 0.62, 0.2)   & $l_2$ \\ 
2 & (1, 1, 0) & (1.77, 0.93, 0.3) & (0.77, -0.07, 0.3)  & $l_1$\\ 
3 & (2, 1, 0) & (2.36, 1.24, 0.4) & (0.36, 0.24, 0.4)   & $l_3$ \\ 
4 & (2, 1, 1) & (2.95, 1.55, 0.5) & (0.95, 0.55, -0.5)  & $l_1$ \\ 
5 & (3, 1, 1) & (3.54, 1.86, 0.6) & (0.54, 0.86, -0.4)  & $l_2$\\ 
6 & (3, 2, 1) & (4.13, 2.17, 0.7) & (1.13, 0.17, -0.3)  & $l_1$\\ 
7 & (4, 2, 1) & (4.72, 2.48, 0.8) & (0.72, 0.48, -0.2)  & $l_1$\\ 
8 & (5, 2, 1) & (5.31, 2.79, 0.9) & (0.31, 0.79, -0.1)  & $l_2$\\ 
9 & (5, 3, 1) & (5.9, 3.1, 1.0)   & (0.9, 0.1, 0.0)     & $l_1$\\ \hline 
10 & (6, 3, 1) &  &  & \\ \hline
\end{tabular}
\label{tab:prop} 
\end{table*}

To alleviate these concerns we use {\em deterministic proportional
routing}.  In this scheme, for each destination, each routing agent
keeps track of the number of packets $pkt_i$ sent though each outgoing
link $l_i$, along with the total number of packets sent
$pkt_{total}$. Deterministic proportional routing consists of sending
packets down the link which has the largest discrepancy between the
desired proportion of packets sent though that link ($p_i *
pkt_{total}$) and the actual number of packets sent through that link
($pk_i$).

Let's consider the following example to illustrate this method: A
routing agent has three outgoing links, $l_1$, $l_2$, and $l_3$,
and its current proportion vector is ($0.59$, $0.31$, $0.1$). If
this agent needs to send 10 packets before changing its proportion
vector, it should send 6, 3 and 1 packets respectively along each of
the outbound links.

Table~\ref{tab:prop} shows how each successive routing decision is
made in this situation.  The first column has the total number of
packets that have been routed, while the second column details the
cumulative number of packets that have been sent down each outgoing
link. The third column shows the ``desired'' packet split at this
instance, which is formed by multiplying the total number of packets
by the proportion vector. (Note that since this will in general
provide fractional packets, it cannot be the actual split.)  The
fourth column shows the difference between the actual split and the
desired split. Finally, the last column gives gives the largest entry
of the fourth column, which is the link to which the next packet
should be sent.

As the splits indicate, this online method not only routes packets in
a way that results in the optimal split over all 10 packets, but also
selects the best split at each intermediate step. (Formally, one can
prove this optimality holds for a large suites of metrics measuring
how bad a particular discrepancy between desired and actual vectors
is, including in particular the $L^2$ and $L^1$ metrics.)

As mentioned in the introduction, it is possible to
implement this scheme extremely quickly, using only additions and
pairwise comparisons. In addition, the scheme can be modified to allow
more recent routing decisions to matter more than older ones, to
prevent ``round-robining'' in which packets arrive out of order at
their ultimate destination, etc. See ~\cite{wolp99a}.


\section{Learning Base Proportions}
\label{sec:learn}
The focus of our study wasn't on finding optimal RL algorithms for
routing, but rather on determining whether RL-based agents running in
an adaptive agent interaction structure could outperform conventional
routing algorithms. Accordingly, the RL algorithms we used were rather
unsophisticated.  All of them bin time into successive {\bf (learning)
intervals}. The actions (i.e., applied proportion vectors) of the
individual routing agents are not allowed to change across a learning
interval. These intervals serve as the smallest observable time unit
for the generation of learning data.  Accordingly, they need to be
long enough to obtain an unambiguous estimate of what system-wide
throughput would be if the actions currently being undertaken by the
agents were continued indefinitely, i.e., long enough to allow for the
current proportion vectors to dominate any lingering effects from the
previous set of proportion vectors. Conversely, we do not want the
interval to be too long, lest it take too long to generate training
data, and more generally to allow the agents to adapt to changes to
network traffic.

\subsection{Learning Algorithms}
\label{sec:steps}
All of the RL algorithms we investigated involved the following
three successive stages:
\begin{enumerate}
\item {\bf Initialization:}
The agents ascertain the network topology. This is a conventional
stage needed for any network to ``boot up''.
\item {\bf Training:}
The agents explore the action space to collect data that will be
subsequently used by the learners. A fixed sequence of different
proportion vectors are applied by the set of all routing agents and
the associated sequence of system-wide throughputs for all those learning
intervals is recorded.  Each element of this sequence will generate an
RL input/output pair for each agent. For each agent, for each
interval, the ``input'' is the action taken 
by the agent together with any features concerning the
network (e.g., proportion vectors of other agents) it observes during
the associated interval, and the output is the system-wide throughput
for that interval. This stage can be viewed either as part of the boot
process of the system, which generates the initial training sets for the agents,
or as a way of mimicking behavior in the middle of an ongoing system by
forming a rough guess for the ``mid-stream'' training sets in that
system.
\item {\bf Learning:}
Choose actions and thereby generate more training input/output pairs,
trading off exploration and exploitation as one does so:
\begin{itemize}
\item
Immediately after the training stage, for each learning agent and for each
destination:
\begin{itemize}
\item
Sweep through the possible proportion values (range of actions),
ranging from 0 to 1.0 in increments of .05. (In our experiments, $m$
was always 2, so proportion vectors reduced to single-dimensional real numbers.)
\item
For each such point, find the $k$ nearest neighbors in the training
set (nearest in the input space), and use these neighbors to estimate
the corresponding system-wide throughput with a memory based learning algorithm.
(Examples of such a learning algorithm are taking a simple mean of those $k$
throughputs as one's estimate, or forming a LMS linear fit through those
$k$ points).
\item
Select the point with the best estimated system-wide throughput and
set the proportion vector to this value for the duration of the
current learning interval. 
\end{itemize}
\item
For subsequent learning intervals:
\begin{itemize}
\item
Store the input/output example generated in the previous learning
interval. 
\item
Sample $n$ values near the previous proportion vector by sampling a
Gaussian centered there.
\item
For each such sample point, find its $k$ nearest neighbors
in the training set and use these neighbors to estimate the
corresponding system-wide throughput as above.
\item
Use a Boltzmann distribution
\footnote{A Boltzmann distribution allows one to balance exploration
against exploitation so that one doesn't get stuck in suboptimal parts
of the space. It does this by selecting actions in a probabilistic
manner, where the actions with the higher immediate payoffs have a
larger probability of being selected. The temperature parameter of the
distribution determines the amount of exploration performed (a low
temperature means that the best action has a high probability of being
selected, whereas a high temperature means that actions have a more
even likelihood of being selected). See ~\cite{wowh99a}.}
over those $n$ estimated throughputs to
select a proportion vector to be applied for the duration of the
current learning interval. All three of $k$, $n$, and the Boltzmann
temperature are held fixed throughout the run.  
\end{itemize}
\end{itemize}
\end{enumerate}


\subsection{Leader-Follower Learning}
\label{sec:leader}

In this adaptive agent interaction structure some agents have empty
feature spaces. Other agents include the actions of some of the 
agents of the first type in their input spaces. Agents of this second
type are called {\bf followers}, and the agents whose actions
followers include in their input spaces are called {\bf leaders}. Note
that to apply this scheme, at the beginning of each learning interval
the leaders must first decide on their actions, and then the followers
use those choices to decide on their actions.

\subsection{Interleaved Learning}
\label{sec:inter}

One potentially unsatisfactory aspect of the leader-follower structure
is the asymmetric way it breaks down the agents. A less asymmetric
adaptive structure has the agents ``interleave'' their decisions in
such a way that every agent is both a leader in some respects, and a
follower in others.  In this scheme, all agents have the actions of
other agents in their feature spaces. However the agents are broken
into two separate groups, where the learning intervals of the two
groups are offset from each other. The offset is arranged so that
any learning interval for the first group overlaps in equal halves with
two successive learning intervals for the second group, and
vice-versa. (In other words, if the intervals of the first group extend
from times 1 to 3, 3 to 5, etc., those of the second group extend from 2
to 4, 4 to 6, etc.)

\section{Experimental Results}
\label{sec:results}
A series of experiments was conducted using the OpNet discrete event
network simulator (version 4.04). Each router in OpNet has an
inbound queue and an outbound queue. Links between routers have infinite
speed but limited bandwidth (1000 bits / simulated second). The Bellman-Ford
algorithm utilized in the experiments was the implementation included
with OpNet.

For the experiments
in this article, this time delay was experimentally determined to be
500 time units on the network discussed in
Section~\ref{sec:network}. 
For the exploration step in the learning interval discussed in 
Section~\ref{sec:steps}, a Gaussian centered on the current proportion
vector with a standard deviation of .0025 was used to generate
5 new proportion vectors. 
The Boltzmann temperature used to select over those new proportion 
vectors was 3000.

\subsection{Network Description}
\label{sec:network}

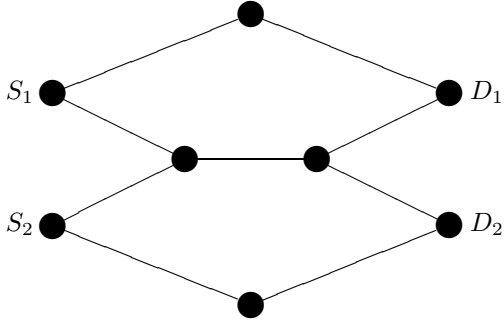
\begin{figure}[bh]
\begin{picture}(100,120)(-40,20)
\put(0,50){\circle*{10}} 
     \put (-8,50) {\makebox(1,1)[r]{$S_2$}}
  \put(0,50) {\line (2,1){50}} 
  \put(0,50) {\line (5,-2){70}} 
\put(0,100){\circle*{10}} 
     \put (-8,100) {\makebox(1,1)[r]{$S_1$}}
  \put(0,100) {\line (2,-1){50}} 
  \put(0,100) {\line (5,2){70}} 

\put(50,75){\circle*{10}} 
     \put (50,75) {\makebox(-1,1)[b]{}}
  \put(50,75) {\line (1,0){50}} 
\put(100,75){\circle*{10}} 
     \put (100,75) {\makebox(-1,1)[b]{}}
  \put(100,75) {\line (2,1){50}} 
  \put(100,75) {\line (2,-1){50}} 

\put(75,20){\circle*{10}} 
     \put (75,10) {\makebox(-1,1)[b]{}}
  \put(75,20) {\line (5,2){70}} 
\put(75,130){\circle*{10}} 
     \put (75,140) {\makebox(-1,1)[t]{}}
  \put(75,130) {\line (5,-2){70}} 

\put(150,50){\circle*{10}} 
     \put (158,50) {\makebox(1,1)[l]{$D_2$}}
\put(150,100){\circle*{10}} 
     \put (158,100) {\makebox(1,1)[l]{$D_1$}}

\end{picture}
\caption{The GEMINI network}
\label{fig:gemini}

\end{figure}

The ``Gemini'' network shown in Figure~\ref{fig:gemini} was used for
testing the various routing approaches. In our experiments, routers
$S_1$ and $S_2$ are the sources where all packets are generated. Nodes
$D_1$ and $D_2$ are the possible (ultimate) destination nodes. Packets
generated at $S_i$ are sent to $D_i$ (for $i=1,2$). The traffic stream
was simulated by generating packets (consisting of 1000 bits each) at
both source nodes with the time between successive packets determined
by randomly sampling the intervals $[.24s,.26s]$ and $[.28s,.30s]$,
respectively.  The intermediate (non-source) routers have their
proportion vectors set to direct 90\% of their packets forward toward
the appropriate destination nodes.

\subsection{Experimental Setup}

The performance of each routing method was evaluated over 40
trials. The initialization stage was $20s$ long, the training stage
then ran from time $t=20s$ to $t=15,000s$, and the learning stage from
$t=15,000s$ to $t=40,000s$. The total delay was measured as the sum of
the delays of all packets generated during the learning period. The
simulation continued for $5000s$ beyond the learning stage to allow
the packets generated in the learning stage to reach their
destinations. During that time the source nodes continued to generate
new packets in order to maintain stationary conditions for the packets
in transit.  Those packets generated during this time were not
included in the calculation of total delay however.

\begin{table*}[htb] \centering
\caption{Performance of baseline and agent-based methods for the Gemini network.
}
\vspace{.1in}
\begin{tabular}{c|c|r|r} \hline
Algorithm & Learning method     &  Total Cost & Error Bar 
\\ \hline \hline
Bellman-Ford & N/A & 1,537,914 & $\pm$2,649 \\ 
BF source only & N/A & 1,180,252 & $\pm$1,520 \\
Ideal Proportions & N/A & 444,637 & $\pm$1,329 \\ \hline
STD-MBL & MBL & 519,557 & $\pm$10,274 \\
LF-MBL & Leader/Follower & 500,339 & $\pm$6,767 \\
INT-MBL & Interleaving & 500,065 & $\pm$7,465 \\ \hline
\end{tabular}
\label{tab:results}
\end{table*}


\subsection{Results and Analysis}

The results of the experiments are summarized in
Table~\ref{tab:results}. The first row contains the performance of BF.
The row labeled ``BF source only'' reports performance when the source
nodes used BF and the intermediate nodes operated with soft masked
proportional routing. This allows a direct comparison of the effects
of replacing BF with memory-based learning methods at the source
nodes. The third row of the table summarizes performance when the
ideal proportion vector is used. (Those vectors were ascertained by
exhaustively running through a suite of simulations in each of which
the proportion vector never changed in the ``learning period''.) Under
the rough assumption that these numbers constitute an upper bound on
performance with the learners, we can use these numbers to provide us
with the ``headroom'' of each algorithm, that is with the amount by
which each algorithm's performance falls short of the best possible
performance.

The algorithms used by the RL-based methods are presented next. The
type of fit was linear based on the 12 nearest neighbors. The learners
reduced the performance headroom between Bellman-Ford and using the
ideal proportions by 93-95\%. Clearly, the agent-based approaches
benefit from using the more sophisticated throughput
estimates. Comparing the agent-based approaches to one another,
leader-follower and interleaved learning reduce the performance
headroom between the standard learner and the ideal proportions by
25\%. Thus, the agent-based approaches where one
or both of the learners have knowledge of the intentions/actions of
the other agent have significantly better performance than the
standard learner.

\section{Discussion}

Adaptivity is a feature of a MAS that becomes increasingly important
the larger the MAS and the less reliable the environment in which it
operates. Broadly speaking, adaptivity takes two forms: adaptivity of
the individual agents, and adaptivity of the interaction structure
among the agents. We have investigated both forms of adaptivity in the
important context of routing over networks. In a set of experiments we
found that simply modifying the action spaces of the agents to make
them better suited to adaptive algorithms potentially improved
throughput by up to a factor of 3.5 over the traditional Bellman-Ford
algorithm. We then investigated two schemes for how to have the agent
interaction structure itself be adaptive. We found that these schemes
both realized a significant fraction of this potential improvement,
with an improvement factor of 3 over Bellman-Ford. Furthermore, a 25\%
improvement was observed over an agent-based approach with no adaptive
interaction structure.

\section{Acknowledgements}
The authors would like to thank Jeremy Frank, Joe Sill,  Ann Bell and
Marjory Johnson for helpful discussion.

\bibliographystyle{plain}

\begin{thebibliography}{10}

\bibitem{ahma93}
R.~K. Ahuja, T.~L. Magnanti, and J.~B. Orlin.
\newblock {\em Network Flows}.
\newblock Prentice Hall, New Jersey, 1993.

\bibitem{baol99}
T.~Basar and G.J. Olsder.
\newblock {\em Dynamic Noncooperative Game Theory}.
\newblock Siam, Philadelphia, PA, 1999.
\newblock Second Edition.

\bibitem{bell57}
R.~E. Bellman.
\newblock {\em Dynamic Programming}.
\newblock Princeton University Press, Princeton, NJ, 1957.

\bibitem{bell58}
R.~E. Bellman.
\newblock On a routing problem.
\newblock {\em Quarterly of applied mathematics}, 16:87--90, 1958.

\bibitem{bega92}
D.~Bertsekas and R.~Gallager.
\newblock {\em Data Networks}.
\newblock Prentice Hall, Englewood Cliffs, NJ, 1992.

\bibitem{boli94}
J.~Boyan and M.~Littman.
\newblock Packet routing in dynamically changing networks: A reinforcement
  learning approach.
\newblock In {\em Advances in Neural Information Processing Systems - 6}, pages
  671--678. Morgan Kaufmann, 1994.

\bibitem{clbo98}
C.~Claus and C.~Boutilier.
\newblock The dynamics of reinforcement learning cooperative multiagent
  systems.
\newblock In {\em Proceedings of the Fifteenth National Conference on
  Artificial Intelligence}, pages 746--752, June 1998.

\bibitem{depa84}
N.~Deo and C.~Pang.
\newblock Shortest path algorithms: Taxonomy and annotation.
\newblock {\em Networks}, 14:275--323, 1984.

\bibitem{dijk59}
E.~Dijkstra.
\newblock A note on two problems in connection with graphs.
\newblock {\em Numeriche Mathematics}, 1(269-171), 1959.

\bibitem{fofu56}
L.~R. Ford and D.~R. Fulkerson.
\newblock Maximal flow through a network.
\newblock {\em Canadian Journal of Mathematics}, 8:419--433, 1956.

\bibitem{goro99}
C.~V. Goldman and J.~S. Rosenschein.
\newblock Emergent coordination through the use of cooperative state--changing
  rules.
\newblock (pre-print), 1999.

\bibitem{hesn98}
M.~Heusse, D.~Snyers, S.~Guerin, and P.~Kuntz.
\newblock Adaptive agent-drivent routing and load balancing in communication
  networks.
\newblock {\em Advances in Complex Systems}, 1:237--254, 1998.

\bibitem{huwe98b}
J.~Hu and M.~P. Wellman.
\newblock Multiagent reinforcement learning: Theoretical framework and an
  algorithm.
\newblock In {\em Proceedings of the Fifteenth International Conference on
  Machine Learning}, pages 242--250, June 1998.

\bibitem{jesy98}
N.~R. Jennings, K.~Sycara, and M.~Wooldridge.
\newblock A roadmap of agent research and development.
\newblock {\em Autonomous Agents and Multi-Agent Systems}, 1:7--38, 1998.

\bibitem{kola97}
Y.~A. Korilis, A.~A. Lazar, and A.~Orda.
\newblock Achieving network optima using {S}tackelberg routing strategies.
\newblock {\em IEEE/ACM Transactions on Networking}, 5(1):161--173, 1997.

\bibitem{kola95}
Y.~A. Korilis, A.~A. Lazar, and A.~Orda.
\newblock Architechting noncooperative networks.
\newblock {\em IEEE Jl. on Sel. Areas in Communications}, 13(8), 1999.

\bibitem{kumi97}
S.~Kumar and R.~Miikkulainen.
\newblock Dual reinforcement {Q}-routing: An on-line adaptive routing
  algorithm.
\newblock In {\em Artificial Neural Networks in Engineering}, volume~7, pages
  231--238. ASME Press, 1997.

\bibitem{laor97}
A.~A. Lazar, A.~Orda, and D.~E. Pendarakis.
\newblock Capacity allocation under nooncooperative routing.
\newblock {\em IEEE Tran. on Networking}, 5(6):861--871, 1997.

\bibitem{sale97}
T.~Sandholm and V.~R. Lesser.
\newblock Coalitions among computationally bounded agents.
\newblock {\em Artificial Intelligence}, 94:99--137, 1997.

\bibitem{sen97}
S.~Sen.
\newblock {\em Multi-Agent Learning: Papers from the 1997 {AAAI} Workshop
  (Technical Report {WS}-97-03}.
\newblock AAAI Press, Menlo Park, CA, 1997.

\bibitem{sudr97}
D.~Subramanian, P.~Druschel, and J.~Chen.
\newblock Ants and reinforcement learning: A case study in routing in dynamic
  networks.
\newblock In {\em Proceedings of the Fifteenth International Conference on
  Artificial Intelligence}, pages 832--838, 1997.

\bibitem{suba98}
R.~S. Sutton and A.~G. Barto.
\newblock {\em Reinforcement Learning: An Introduction}.
\newblock MIT Press, Cambridge, MA, 1998.

\bibitem{syca98}
K.~Sycara.
\newblock Multiagent systems.
\newblock {\em AI Magazine}, 19(2):79--92, 1998.

\bibitem{tuwo99}
K.~Tumer and D.~H. Wolpert.
\newblock Avoiding {B}raess' paradox through collective intelligence.
\newblock (in preparation), 1999.

\bibitem{wada92}
C.~Watkins and P.~Dayan.
\newblock Q-learning.
\newblock {\em Machine Learning}, 8(3/4):279--292, 1992.

\bibitem{wolp99a}
D.~H. Wolpert.
\newblock Masked proportional routing.
\newblock Patent pending, 1999.

\bibitem{wotu99b}
D.~H. Wolpert and K.~Tumer.
\newblock An {I}ntroduction to {C}ollective {I}ntelligence.
\newblock In J.~M. Bradshaw, editor, {\em Handbook of Agent technology}. AAAI
  Press/MIT Press, 1999.
\newblock to appear; preprint cs.LG/9908014 at xxx.lanl.gov archive.

\bibitem{wotu99a}
D.~H. Wolpert, K.~Tumer, and J.~Frank.
\newblock Using collective intelligence to route internet traffic.
\newblock In {\em Advances in Neural Information Processing Systems - 11},
  pages 952--958. MIT Press, 1999.

\bibitem{wowh99a}
D.~H. Wolpert, K.~Wheeler, and K.~Tumer.
\newblock General principles of learning-based multi-agent systems.
\newblock In {\em Proceedings of the Third International Conference of
  Autonomous Agents}, pages 77--83, 1999.

\end{thebibliography}

\end{document}